\documentclass[superscriptaddress,amssymb,amsmath,floatfix,citeautoscript,twocolumn]{revtex4-1}
\usepackage[pdftex]{graphicx}
\usepackage{chngcntr}
\usepackage{amsmath}
\usepackage{color}
\usepackage{natbib}
\usepackage{mathtools}
\usepackage{physics}
\usepackage{lipsum}

\begin{document}

\title{Theory of the Little-Parks effect in spin-triplet superconductors}


\author{Chengyun Hua}
\email{
E-mail: huac@ornl.gov }
\affiliation{Materials Science and Technology Division, Oak Ridge National Laboratory, Oak Ridge, TN 37831, USA}
\author{Eugene Dumitrescu}
\affiliation{Computational Sciences and Engineering Division, Oak Ridge National Laboratory, Oak Ridge, TN 37831, USA}
\author{G\'abor B. Hal\'asz}
\email{halaszg@ornl.gov}
\affiliation{Materials Science and Technology Division, Oak Ridge National Laboratory, Oak Ridge, TN 37831, USA}

\date{\today}
\begin{abstract}

The celebrated Little-Parks effect in mesoscopic superconducting rings has recently gained great attention due to its potential to probe half-quantum vortices in spin-triplet superconductors. However, despite the large number of works reporting anomalous Little-Parks measurements attributed to unconventional superconductivity, the general signatures of spin-triplet pairing in the Little-Parks effect have not yet been systematically investigated. Here we use Ginzburg-Landau theory to study the Little-Parks effect in a spin-triplet superconducting ring that supports half-quantum vortices; we calculate the field-induced Little-Parks oscillations of both the critical temperature itself and the residual resistance resulting from thermal vortex tunneling below the critical temperature. We observe two separate critical temperatures with a single-spin superconducting state in between and find that, due to the existence of half-quantum vortices, each minimum in the upper critical temperature splits into two minima for the lower critical temperature. From a rigorous calculation of the residual resistance, we confirm that these two minima in the lower critical temperature translate into two maxima in the residual resistance below and establish the general conditions under which the two maxima can be practically resolved. In particular, we identify a fundamental trade-off between sharpening each maximum and keeping the overall magnitude of the resistance large. Our results will guide experimental efforts in designing mesoscopic ring geometries for probing half-quantum vortices in spin-triplet candidate materials on the device scale.

\end{abstract}

\maketitle

\section{Introduction}

Topological quantum computation based on Majorana bound states is a leading candidate for processing quantum information~\cite{nayak_non-abelian_2008}. The vortex cores of topological superconductors, such as gapped $p$-wave superconductors, host such self-conjugate Majorana bound states at zero energy~\cite{read_paired_2000, ivanov_non-abelian_2001, alicea_new_2012}. The search for these unconventional superconductors has greatly intensified in the past few years as new candidate $p$-wave pairing states are proposed both intrinsically in bulk superconductors~\cite{maeno_superconductivity_1994, maeno_intriguing_2001, mackenzie_superconductivity_2003, mackenzie_even_2017, ran_nearly_2019, li_observation_2019, xu_spin-triplet_2020, aoki_unconventional_2022} and on the surfaces of more conventional superconductors~\cite{fu_superconducting_2008, wang_topological_2015, wu_topological_2016, xu_topological_2016, zhang_observation_2018, machida_zero-energy_2019}. 

The Little-Parks effect~\cite{little_observation_1962} originates from the macroscopic quantum coherence of Cooper pairs; due to the quantization of the fluxoid, the resistance of a thin superconducting ring oscillates as a function of the applied magnetic flux. For a conventional $s$-wave superconductor, the periodicity of these Little-Parks oscillations is given by the flux quantum $\Phi_0 = h/2e$, and the minima of the resistance correspond to integer multiples of $\Phi_0$. It has been recognized, however, that unconventional superconductors may exhibit different kinds of Little-Parks oscillations. For example, in gapless superconductors with $d$-wave pairing, the Little-Parks oscillations acquire an enlarged periodicity 2$\Phi_0$~\cite{loder_magnetic_2008,juricic_restoration_2008,zhu_magnetic_2010}, while polycrystalline $p$-wave superconductors have shifted Little-Parks oscillations with the minima of the resistance corresponding to half-integer multiples of $\Phi_0$~\cite{geshkenbein_vortices_1987,xu_spin-triplet_2020,li_observation_2019}. Fractional Little-Parks oscillations with reduced periodicities $\Phi_0 / n$ have also been recently reported both experimentally~\cite{ge_discovery_2022} and theoretically~\cite{halasz_fractional_2021, zhang_higgs-leggett_2022, pan_frustrated_2022}.

For intrinsic spin-triplet $p$-wave superconductors, Majorana bound states have been predicted to emerge in the cores of half-quantum vortices (HQVs) at which half-integer flux quanta $\Phi_0/2$ pierce through the superconductor~\cite{read_paired_2000, ivanov_non-abelian_2001}. In the presence of such HQVs, the Little-Parks oscillations are then expected to possess a distinctive two-peak structure with minima of the resistance at both integer and half-integer multiples of $\Phi_0$~\cite{vakaryuk_effect_2011, yasui_little-parks_2017, cai_magnetoresistance_2022}. At each half-integer minimum, the fluxoid of the superconducting ring is quantized to a half-integer multiple of $\Phi_0$ (meaning that an HQV is bound to the central hole of the ring), while the two peaks around such a minimum correspond to transitions between integer and half-integer fluxoid quantizations. Still, even though this two-peak structure in the Little-Parks oscillations may prove crucial for identifying spin-triplet superconductors, the precise conditions required for its observation are yet to be firmly established. More generally, a rigorous theoretical understanding of the spin-triplet Little-Parks effect could reveal additional signatures of spin-triplet superconductivity and hence provide alternative avenues for detecting this exotic superconducting state on the device scale.

In this work, we employ the Ginzburg-Landau approach to theoretically study the Little-Parks effect in spin-triplet superconducting rings supporting HQVs. We consider both the ``conventional'' Little-Parks oscillations of the critical temperature~\cite{little_observation_1962} and the analogous magnetoresistance oscillations below the critical temperature that result from thermal vortex tunneling~\cite{cai_magnetoresistance_2022, sochnikov_large_2010, sochnikov_oscillatory_2010, cai_unconventional_2013, mills_vortex_2015}. In computing the residual resistance of the superconducting ring below the critical temperature, we do not only focus on the fluxoid ground state~\cite{cai_magnetoresistance_2022, aoyama_little-parks_2022} but also account for the thermally occupied excited states and the thermally activated transitions between them.

We first demonstrate that HQVs are stabilized by an appropriate higher-order term in the Ginzburg-Landau free energy which penalizes the charge supercurrent but not the spin supercurrent. By analyzing the effect of this term on the free energies of the various fluxoid states below the critical temperature, we confirm the presence of a two-peak structure in the magnetoresistance oscillations~\cite{vakaryuk_effect_2011, yasui_little-parks_2017, cai_magnetoresistance_2022} and understand how the separation between the two peaks depends on the temperature. Next, we compute the magnetoresistance oscillations themselves and explicitly quantify the prominence of the characteristic two-peak structure. By identifying a fundamental trade-off between minimizing the width of each peak and maximizing the overall magnitude of the resistance, we provide detailed experimental guidelines for probing the two-peak structure in real candidate materials.

Turning our attention to the ``conventional'' Little-Parks oscillations, we observe two separate critical temperatures with the higher one marking the onset of superconductivity and the lower one separating a single-spin superconducting state above and a spin-triplet superconducting state below. While the two-peak structure is entirely absent from the Little-Parks oscillations of the upper critical temperature, it translates into a two-valley structure for the lower critical temperature, thus providing a further signature of spin-triplet superconductors supporting HQVs.

\section{Ginzburg-Landau theory for a spin-triplet superconductor}

We consider a spin-triplet superconductor with $p_x+ip_y$ pairing symmetry in which spin-orbit coupling energetically favors $(\uparrow\uparrow)$ and $(\downarrow\downarrow)$ Cooper pairs over $(\uparrow\downarrow) + (\downarrow\uparrow)$ Cooper pairs. Such a superconductor can support HQVs around which the superconducting phase of only one type of Cooper pair [either $(\uparrow\uparrow)$ or $(\downarrow\downarrow)$] winds by $2\pi$. Under a magnetic field parallel to the spin quantization axis, the Ginzburg-Landau free energy of such a superconductor is given by~\cite{machida_incomplete_1978}
\begin{widetext}
\begin{eqnarray}\label{eq:FreeEnergy}
F = \frac{T_c F_0} {V_0} \int d^3r\Bigg\{ \sum_{\sigma = \uparrow,\downarrow}\biggl[-\Big\{(1-t)&+&z_\sigma |\grad \times \vec{a}|\Big\}|\psi_\sigma|^2+\frac{|\psi_\sigma|^4}{2}+\xi_0^2\big|(\grad-i\vec{a})\psi_\sigma\big|^2\biggl] \nonumber \\
&+& c|\psi_\uparrow|^2|\psi_\downarrow|^2+\xi_0^2 d \big|(\grad-2i\vec{a})[\psi_\uparrow\psi_\downarrow]\big|^2 +\frac{V_0} {2\mu_0 T_c F_0}(\grad \times \vec{a})^2\Bigg\},
\end{eqnarray}
\end{widetext}
where $t = T/T_c$ is the dimensionless temperature (with $T_c$ being the critical temperature), $\xi_0$ is the zero-temperature coherence length, $\vec{a} = 2\pi \vec{A}/\Phi_0$ is the magnetic vector potential, while $\psi_\uparrow$ and $\psi_\downarrow$ are the superconducting order parameters corresponding to the $(\uparrow\uparrow)$ and $(\downarrow\downarrow)$ Cooper pairs, respectively. The $z_\sigma$ term arises from Zeeman splitting in a magnetic field with $z \equiv z_\uparrow = -z_\downarrow$, the $c$ and $d$ terms  describe coupling between the two types of Cooper pairs, and the last term of Eq.~(\ref{eq:FreeEnergy}) accounts for the magnetic screening effect of the charge supercurrent. We point out that the standard form of the Ginzburg-Landau free energy~\cite{machida_incomplete_1978} only contains terms in which the total number of order parameters $\psi_\sigma$ and spatial derivatives $\grad$ does not exceed $4$. In this work, we include an additional symmetry-allowed term proportional to $d$ with $4$ order parameters and $2$ spatial derivatives that penalizes the charge supercurrent but not the spin supercurrent. As we will later find, this additional term is crucial for stabilizing HQVs. We also note that the dimensionless energy parameter $F_0$ is chosen such that $T_c F_0$ is the condensation energy of the entire superconductor with volume $V_0$ at zero temperature in the absence of a magnetic field ($\vec{a} = 0$) and any coupling ($c=d=0$).

\begin{figure}
\centering
\includegraphics[scale = 0.35]{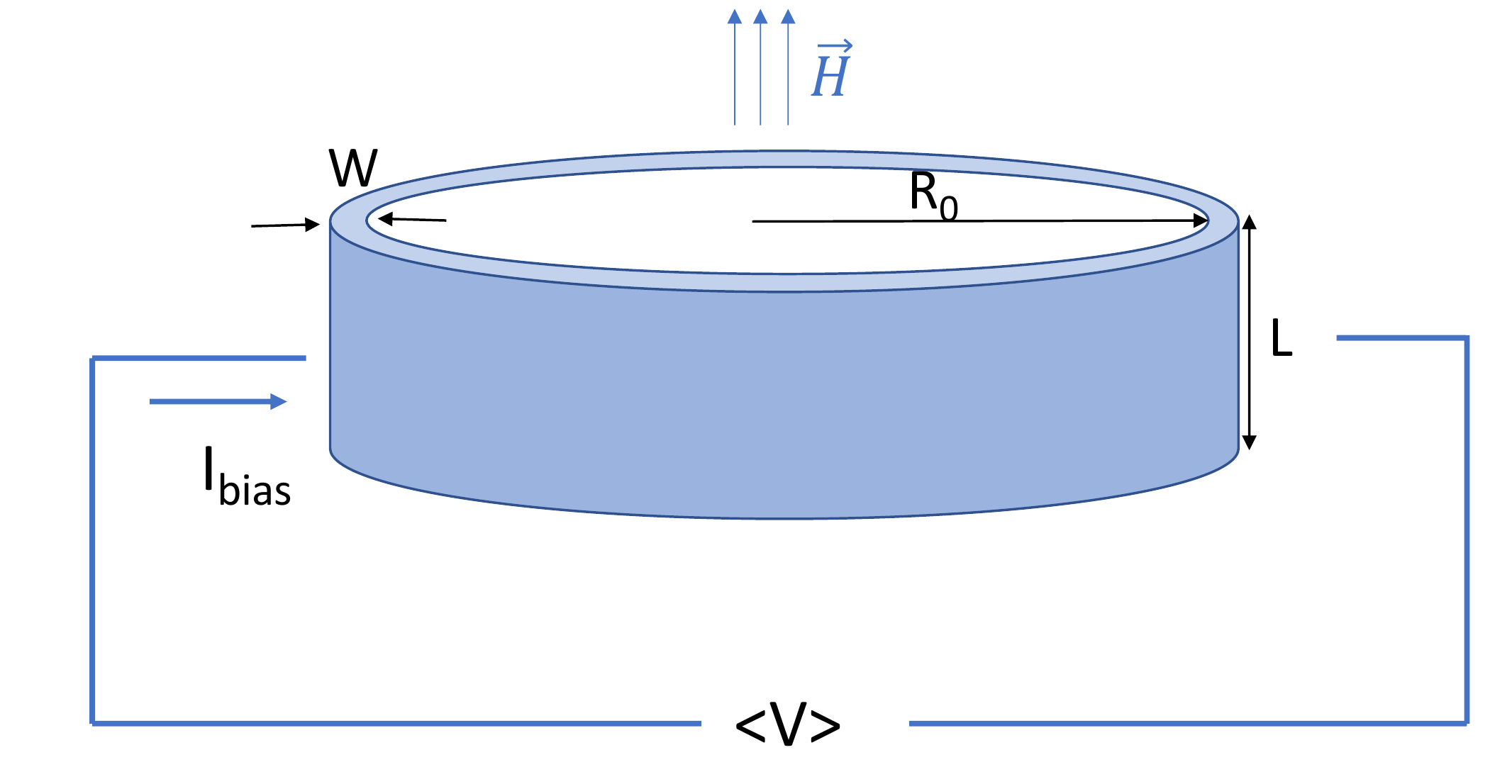}
\caption{Schematics of the Little-Parks experiment. The resistance of a thin superconducting ring with radius $R_0$, width $W \ll R_0$, and height $L$ is measured as a function of the applied magnetic field $\vec{H}$ near the superconducting critical temperature. The resistance itself is determined by applying a bias current $I_{\mathrm{bias}}$ and measuring the resulting voltage $\langle V \rangle$.}
\label{fig:Schematics}
\end{figure} 

In the rest of this work, we focus on a thin superconducting ring of radius $R_0$ and width $W \ll R_0$ in a perpendicularly applied magnetic field $H$ (schematics shown in Fig.~\ref{fig:Schematics}). For $W \ll \xi$, where $\xi = \xi_0 (1-t)^{-1/2}$ is the finite-temperature coherence length, the order parameters $\psi_\sigma$ only depend on the polar angle $\theta$. Moreover, if $W$ is much smaller than the penetration depth $\lambda$, the screening effect of the charge supercurrent is negligible, and the vector potential in the symmetric gauge is simply given by $\vec{A} = \frac{1}{2} R_0 H \hat{\theta}$. Expressing the order parameters as $\psi_\sigma(\theta) = f_\sigma(\theta)e^{i\phi_\sigma(\theta)}$ in terms of the amplitudes $f_\sigma(\theta)$ and phases $\phi_\sigma(\theta)$, the Ginzburg-Landau equations derived from the free-energy functional of Eq.~(\ref{eq:FreeEnergy}) are then
\begin{widetext}
\begin{subequations}\label{eq:GL_LP}
\begin{eqnarray}
&& -\left[(1-t)+\frac{b_\sigma h}{r^2_0}\right]f_\sigma+f^3_\sigma+cf_\sigma f^2_{-\sigma}  = \frac{1}{r^2_0} \left\{ \frac{\partial^2f_\sigma}{\partial\theta^2}-f_\sigma\left[\frac{\partial\phi_\sigma}{\partial\theta}-h\right]^2 \nonumber \right\} \\
&& \qquad \qquad \qquad \qquad \qquad \qquad \qquad \qquad \quad +\frac{d}{r^2_0} \left\{ f_{-\sigma}\frac{\partial^2(f_\sigma f_{-\sigma})}{\partial\theta^2}-f_\sigma f_{-\sigma}^2\left[\frac{\partial(\phi_\sigma+\phi_{-\sigma})}{\partial\theta}-2h\right]^2 \right\}, \label{eq:f_up}\\
&& \frac{\partial}{\partial\theta}\left\{f^2_\sigma\left[\frac{\partial\phi_\sigma}{\partial\theta}-h\right] + d(f_\sigma f_{-\sigma})^2\left[\frac{\partial(\phi_\sigma+\phi_{-\sigma})}{\partial\theta}-2h\right]\right\}= 0. \label{eq:phi_up}
\end{eqnarray}
\end{subequations}
\end{widetext}
Here $r_0 = R_0/\xi_0$ is a dimensionless ring radius, $b_\sigma = 2z_\sigma/\xi_0^2$ is a dimensionless Zeeman splitting, $h = H R_0^2\pi/\Phi_0$ is the number of flux quanta going through the ring, while $-\sigma$ indicates the opposite type of Cooper pair with respect to $\sigma$. We note that the conserved quantity within the curly brackets of Eq.~(\ref{eq:phi_up}) is a supercurrent that corresponds to the given type of Cooper pair [$(\uparrow\uparrow)$ or $(\downarrow\downarrow)$]; the charge and spin supercurrents are symmetric and antisymmetric combinations of these individual supercurrents, respectively. Since the order parameters $\psi_\sigma(\theta)$ must be single valued, the Ginzburg-Landau equations in Eq.~(\ref{eq:GL_LP}) are also supplemented with the boundary conditions $f_\sigma(\pi) = f_\sigma(-\pi)$ and $\phi_\sigma(\pi)-\phi_\sigma(-\pi) = 2\pi n_\sigma$, where $n_\sigma$ are arbitrary integers.  

\section{Field-induced oscillations of the critical temperature}

In this section, we study the ``conventional'' Little-Parks oscillations~\cite{little_observation_1962} in the critical temperature of a spin-triplet superconducting ring as a function of the applied magnetic field. To describe these Little-Parks oscillations, we must establish the ground-state phase diagram of the system by enumerating and comparing ground-state candidates: stable solutions of Eq.~(\ref{eq:GL_LP}) that correspond to local minima of the free-energy functional in Eq.~(\ref{eq:FreeEnergy}). Such stable solutions take the general form of $f_\sigma (\theta) = \mathrm{constant}$ and $\phi_\sigma (\theta) = \phi^{(0)}_\sigma + n_\sigma \theta$, where $n_\sigma \in \mathbb{Z}$ are the fluxoid numbers for the two types of Cooper pairs and $\phi^{(0)}_\sigma$ are arbitrary reference phases~\cite{mccumber_time_1970}. Substituting this form into Eq.~(\ref{eq:f_up}), the constant values of the order parameters $f_\sigma$ are then solutions of the algebraic equations
\begin{eqnarray}\label{eq:solution_1stkind}
f_\uparrow \left[ f^2_\uparrow + \tilde{c} (n_\uparrow,n_\downarrow) f^2_\downarrow - \tilde{\alpha}_\uparrow (n_\uparrow) \right] &=& 0, \nonumber \\
\qquad f_\downarrow \left[ f^2_\downarrow + \tilde{c} (n_\uparrow,n_\downarrow) f^2_\uparrow - \tilde{\alpha}_\downarrow (n_\downarrow) \right] &=& 0,
\end{eqnarray}
where $\tilde{\alpha}_\sigma (n_\sigma) = (1-t)-r^{-2}_0[(n_\sigma-h)^2 - b_\sigma h]$ and $\tilde{c} (n_\uparrow,n_\downarrow) = c+r^{-2}_0 d (n_\uparrow+n_\downarrow -2h)^2$. These equations have four families of solutions that are listed in Table~\ref{tab:solutions} along with their free energies and physical interpretations. The trivial solution with $f_\uparrow = f_\downarrow = 0$ corresponds to a normal state, while the three nontrivial families describe superconducting states. For the $(n_\uparrow)_\uparrow$ and $(n_\downarrow)_\downarrow$ solutions, only one spin species [either $\uparrow$ or $\downarrow$] forms Cooper pairs with the other spin species remaining in a normal state. These solutions correspond to an effectively spinless superconductor that can only support full quantum vortices (FQVs) with a single integer fluxoid number ($n_\uparrow$ or $n_\downarrow$). In contrast, for the $(n_\uparrow,n_\downarrow)$ solutions, both spin species form Cooper pairs, and the result is a spin-triplet superconductor with both $(\uparrow\uparrow)$ and $(\downarrow\downarrow)$ pairing. Introducing the charge $n_c = (n_\uparrow + n_\downarrow)/2$ and spin $n_s = (n_\uparrow - n_\downarrow)/2$ fluxoid numbers, and recognizing that $n_c$ is the ``usual'' fluxoid number connected to the magnetic field, it is then clear that such a spin-triplet superconductor can support both FQVs (integer $n_{c,s}$) and HQVs (half-integer $n_{c,s}$).

\begin{table*}[t]
    \centering
    \renewcommand{\arraystretch}{2}
    \begin{tabular}{|c|c|c|c|c|}
        \hline Solution & $f_\uparrow^2$ & $f_\downarrow^2$ & Free energy ($F$) & Physical interpretation \\
        \hline Trivial & 0 & 0 & 0 & Normal (non-superconducting) state \\
        \hline $(n_\uparrow)_\uparrow$ & $\tilde{\alpha}_\uparrow$ & 0 & $-\frac{1}{2} F_0 \tilde{\alpha}_\uparrow^2$ & ``Spinless'' superconducting state with only ($\uparrow\uparrow$) Cooper pairs \\
        \hline $(n_\downarrow)_\downarrow$ & 0 & $\tilde{\alpha}_\downarrow$ & $-\frac{1}{2} F_0 \tilde{\alpha}_\downarrow^2$ & ``Spinless'' superconducting state with only ($\downarrow\downarrow$) Cooper pairs \\
        \hline $(n_\uparrow, n_\downarrow)$ & $\dfrac{\tilde{\alpha}_\uparrow - \tilde{c} \tilde{\alpha}_\downarrow} {1-\tilde{c}^2}$ & $\dfrac{\tilde{\alpha}_\downarrow - \tilde{c} \tilde{\alpha}_\uparrow} {1-\tilde{c}^2}$ & $-\frac{1}{2} F_0 \dfrac{\tilde{\alpha}_\uparrow^2 + \tilde{\alpha}_\downarrow^2 - 2\tilde{c} \tilde{\alpha}_\uparrow \tilde{\alpha}_\downarrow} {1-\tilde{c}^2}$ & Triplet superconducting state with ($\uparrow\uparrow$) and ($\downarrow\downarrow$) Cooper pairs \\ \hline
    \end{tabular}
    \caption{Four families of solutions for the order parameters $f_\sigma$ along with their respective free energies and physical interpretations. Each non-trivial family contains infinitely many distinct solutions labeled by the fluxoid numbers $n_\sigma$; the solutions for $f_\sigma$ and the corresponding free energies depend on $n_\sigma$ via $\tilde{\alpha}_\sigma (n_\sigma)$ and $\tilde{c} (n_\uparrow, n_\downarrow)$.}
    \label{tab:solutions}
\end{table*}

\begin{figure*}
\centering
\includegraphics[scale = 0.35]{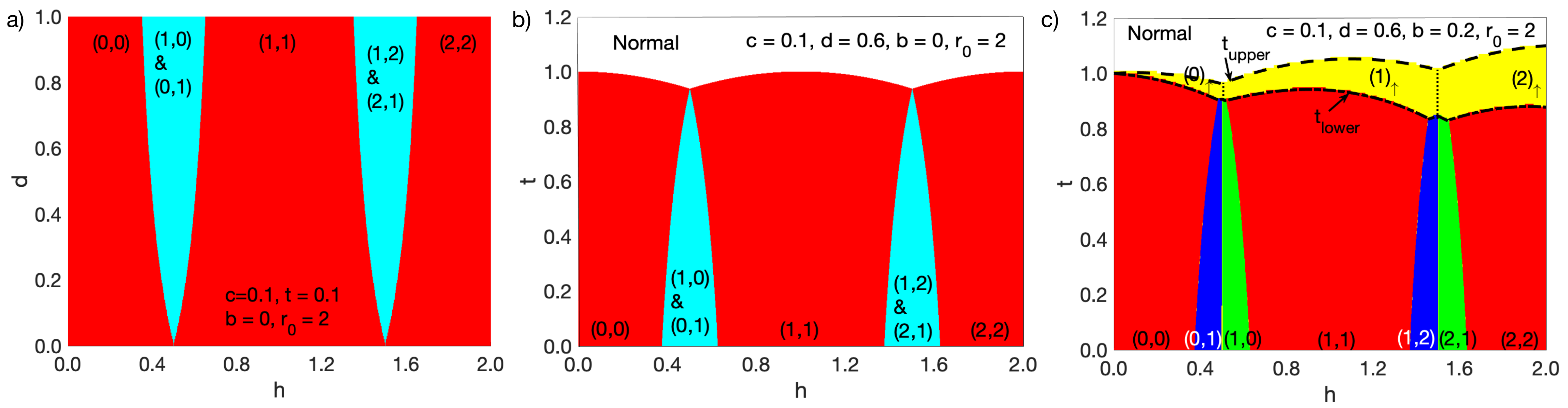}
\caption{Phase diagram of the ground-state fluxoid state (a) as a function of magnetic flux $h$ and coupling coefficient $d$ without Zeeman splitting ($b = 0$), (b) as a function of magnetic flux $h$ and temperature $t$ without Zeeman splitting, and (c) as a function of $h$ and $t$ with Zeeman splitting ($b \neq 0$). The white area marks the normal state, the yellow area marks a single-spin superconducting state with $f^2_\downarrow = 0$, and the red area marks a spin-triplet superconducting state with an integer fluxoid (FQV) around the ring ($n_{c,s} \in \mathbb{Z}$). In (a) \& (b), the light blue area indicates a degeneracy between two half-integer fluxoid (HQV) states with $n_s = \pm 1/2$, while in (c), the green (blue) color means that the HQV state with $n_s = +1/2$ ($n_s = -1/2$) is the ground state. In (c), the dashed and dash-dotted lines mark the upper and lower critical temperatures [Eqs.~(\ref{eq:tc_upper}) \& (\ref{eq:tc_lower})], respectively.}
\label{Figure1}
\end{figure*} 

To determine the ground state of the ring, we need to compare the free energies of all solutions in Table~\ref{tab:solutions} while keeping in mind that each solution is only physical if $f_\uparrow^2 \geq 0$ and$f_\downarrow^2 \geq 0$. The resulting phase diagrams as a function of the temperature $t$, the magnetic field $h$, and the coupling constant $d$ are plotted in Fig.~\ref{Figure1}. We start understanding these results by comparing the various solutions within each family of Table~\ref{tab:solutions}. For the single-spin superconducting states $(n_\uparrow)_\uparrow$ and $(n_\downarrow)_\downarrow$ [marked by yellow color in Fig.~\ref{Figure1}], the free energy takes the exact form
\begin{equation}\label{eq:single-spin}
F_{(n_\sigma)_\sigma} = -\frac{F_0}{2} \left\{ (1-t) - \frac{1}{r^2_0} \left[ (n_\sigma - h)^2 - b_\sigma h \right] \right\}^2.
\end{equation}
We note that the expression inside the curly brackets must be positive for the given state to be valid. Hence, assuming $n \leq h \leq n+1$ (where $n$ is a non-negative integer) without loss of generality, the single-spin state with the lowest free energy at finite Zeeman splitting $b > 0$ (with $b \equiv b_\uparrow = -b_\downarrow$) is $(n)_\uparrow$ for $h < n+1/2$ and $(n+1)_\uparrow$ for $h > n+1/2$. For the spin-triplet states $(n_\uparrow,n_\downarrow)$, we expand the free energy up to $O(r_0^{-2})$ and express $n_\sigma$ in terms of $n_{c,s}$ to obtain
\begin{widetext}
\begin{equation}
F_{(n_\uparrow,n_\downarrow)} = F_0 \biggl\{-\frac{(1-t)^2}{1+c}+\frac{2(1-t)}{r_0^2}\frac{(1+c)n^2_s+[(1+c)+2(1-t)d](n_c-h)^2}{(1+c)^2}+O(r_0^{-4})\biggl\}. \label{eq:FreeEnergy_minima}
\end{equation}
\end{widetext}

Therefore, in the $r_0 \gg 1$ limit, we can restrict our attention to the FQV states with $n_s = 0$ and the HQV states with $n_s = \pm 1/2$ as all other states are penalized by the term $\propto n_s^2$. If we assume $n \leq h \leq n+1$ (with $n \in \mathbb{Z}$ and $n \geq 0$) again, the lowest-energy FQV state has $n_c = n$ for $h < n+1/2$ and $n_c = n+1$ for $h > n+1/2$. These two states correspond to $(n,n)$ and $(n+1,n+1)$ in the notation of Table~\ref{tab:solutions} and are both denoted by red color in Fig.~\ref{Figure1}. The two lowest-energy HQV states with $n_c = n+1/2$ and $n_s = \pm 1/2$, corresponding to $(n+1,n)$ and $(n,n+1)$, are degenerate up to $O(r_0^{-2})$. This degeneracy is split by a higher-order term in the free energy, $4 r_0^{-4} b h n_s (n_c - h) / (1 - c)$, such that the lowest-energy HQV state is $(n,n+1)$ for $h < n+1/2$ [blue color in Fig.~\ref{Figure1}] and $(n+1,n)$ for $h > n+1/2$ [green color in Fig.~\ref{Figure1}]. To compare the lowest-energy FQV and HQV states with one another, we finally recognize that the terms $\propto n_s^2$ and $\propto (n_c-h)^2$ have relative coefficients $(1+c)$ and  $(1+c)+2(1-t)d$ in Eq.~(\ref{eq:FreeEnergy_minima}). Given that $n_s^2 = 1/4$ for the HQV state and $0 \leq (n_c-h)^2 \leq 1/4$ for both states, the HQV state can only have lower energy than the FQV state if the coupling constant $d$ is positive. Thus, as previously stated, this coupling constant is crucial for stabilizing HQV states. Specifically, we find that the lowest-energy spin-triplet state is a HQV state within the field range $n+1/2-\Delta h/2 < h < n+1/2+\Delta h/2$ characterized by the width parameter
\begin{equation}
\Delta h = \frac{(1-t)d} {(1+c)+2(1-t)d} + O(r_0^{-2}). \label{eq:Delta-1}
\end{equation}
As shown by Eq.~(\ref{eq:Delta-1}) as well as Fig.~\ref{Figure1}, the field range in which a HQV state is energetically favorable increases as the temperature $t$ is lowered and as the coupling constant $d$ is increased. Conversely, this field range vanishes both at the critical temperature ($t \to 1$) and when the relevant coupling constant vanishes ($d \to 0$).

We are now ready to understand the field dependence of the critical temperature. We first recognize that, for $b h \neq 0$, there are in fact two critical temperatures; in addition to the upper critical temperature at which superconductivity first appears for one spin species, there is also a lower critical temperature marking a transition from a single-spin superconductor into a spin-triplet superconductor. These two critical temperatures can be determined by comparing the free energies of the lowest-energy single-spin and spin-triplet superconducting states [see Eqs.~(\ref{eq:single-spin}) and (\ref{eq:FreeEnergy_minima})] with each other and with the free energy of the normal state (which is zero). The upper critical temperature is found to be
\begin{equation}
t_{\mathrm{upper}} = 1 + \frac{1}{r^2_0} \left[ b|h| - (h - \lfloor h \rceil)^2 \right], \label{eq:tc_upper}
\end{equation}
where $\lfloor h \rceil$ is simply $h$ rounded to the nearest integer. This critical temperature has the same field-induced oscillation as the critical temperature of a spin-singlet superconductor~\cite{little_observation_1962} except for an additional linear increase with the field $|h|$ that shifts the maxima around integer $h$ [see Fig.~\ref{Figure1}(c)]. Up to $O(r_0^{-4})$, the lower critical temperature is given by
\begin{widetext}
\begin{equation}
t_{\mathrm{lower}} = 1 - \frac{[h - f(h)]^2 - c (h - \lfloor h \rceil)^2 + (1+c) b|h|} {r^2_0 (1-c)}- \frac{d [2h - \lfloor h \rceil - f(h)]^2 \{[h - f(h)]^2 - (h - \lfloor h \rceil)^2 + 2b|h|\}} {r^4_0 (1-c)^2} + O(r_0^{-6}) \label{eq:tc_lower}
\end{equation}
\end{widetext}
in terms of the piece-wise continuous function
\begin{equation}
f(h) = \bigg\{ \begin{matrix} \lfloor h \rceil \qquad \qquad \quad \mathrm{if} \,\, |h - \lfloor h \rceil| < (1 - \Delta h) / 2, \\ \lfloor 2h \rceil - \lfloor h \rceil \qquad \mathrm{if} \,\, |h - \lfloor h \rceil| > (1 - \Delta h) / 2, \end{matrix} \label{eq:fh}
\end{equation}
where the upper and the lower cases correspond to transitions into FQV and HQV states, respectively, and the field ranges around half-integer $h$ with transitions into HQV states are controlled by the width parameter
\begin{equation}
\Delta h = \frac{2db|h|} {r^2_0 (1-c)} + O(r_0^{-4}). \label{eq:Delta-2}
\end{equation}
In contrast to the upper critical temperature, the lower critical temperature shows an overall decrease with the field $|h|$ and contains additional maxima at all half-integer values of $h$ [see Fig.~\ref{Figure1}(c)] around which the transitions into the HQV states happen. We emphasize that, while Eqs.~(\ref{eq:Delta-1}) and (\ref{eq:Delta-2}) describe the same width parameter $\Delta h$, one cannot obtain Eq.~(\ref{eq:Delta-2}) by simply substituting Eq.~(\ref{eq:tc_lower}) into Eq.~(\ref{eq:Delta-1}) because the leading term of Eq.~(\ref{eq:Delta-2}) is $O(r_0^{-2})$ whereas Eq.~(\ref{eq:Delta-1}) is only accurate up to $O(1)$ terms. We further remark that a spin-singlet superconductor with multiple bands does not have two separate critical temperatures because its Ginzburg-Landau free energy includes bilinear coupling terms between different order parameters~\cite{erin_little-parks_2008}. Such coupling terms of the form $\propto \psi_{-\sigma}^* \psi_{\sigma}^{\phantom{*}}$ are forbidden in our case by spin-rotation symmetry around the field direction (i.e., the direction perpendicular to the ring).

\section{Magnetoresistance oscillations below the critical temperature}

In this section, we study the resistance of the superconducting ring in the spin-triplet state below the lower critical temperature in order to derive its oscillations as a function of the applied magnetic field. To do so, we will first consider thermal fluctuations in the order parameters $\psi_\sigma (\theta)$ that drive transitions between different free-energy minima. We will then use the computed transition rates to estimate the electrical resistance due to the resulting thermal decay of the charge supercurrent~\cite{mccumber_time_1970, cooper_resistance_2010}.

\begin{figure*}
\centering
\includegraphics[scale = 0.3]{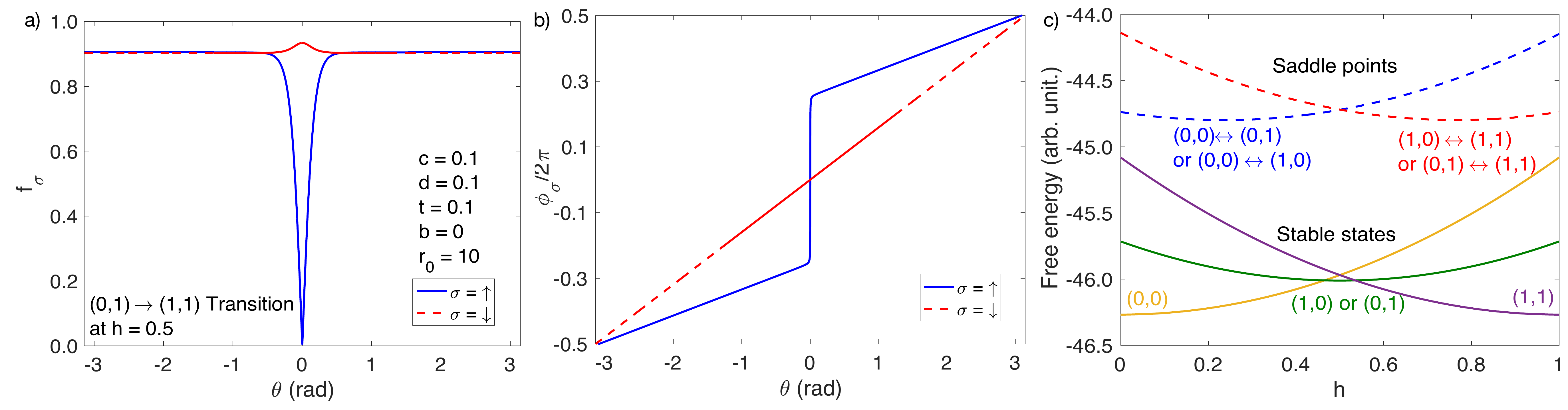}
\caption{(a,b) Saddle-point solutions for (a) the amplitude and (b) the phase of the superconducting order parameters $\psi_\uparrow$ and $\psi_\downarrow$ as the fluxoid numbers $(n_\uparrow,n_\downarrow)$ change from $(0,1)$  to $(1,1)$ at magnetic flux $h = 0.5$. (c) Free energies of the relevant stable states (solid lines) and the saddle points connecting them (dashed lines) as a function of the magnetic flux $h$.}
\label{Figure2}
\end{figure*}

During a thermal transition from one free-energy minimum to another one, the system goes through an appropriate free-energy saddle point, and the free-energy barrier controlling the transition rate is simply the free-energy difference between the saddle point and the original minimum. The two free-energy minima connected by the thermal transition correspond to two stable solutions of Eq.~(\ref{eq:GL_LP}) that differ in a fluxoid number $n_\sigma$. Therefore, the saddle-point solution must exhibit a phase slip in the corresponding order parameter $\psi_\sigma (\theta)$, i.e., a suppression of the amplitude $f_\sigma (\theta)$ around some angle $\theta$. For a spin-singlet superconductor, an analytical expression for this saddle-point solution was found in Ref.~\onlinecite{mccumber_time_1970}. Since the analytical solution only applies for a spin-triplet superconductor in the absence of coupling terms ($c=d=0$), we choose to solve Eq.~(\ref{eq:GL_LP}) numerically with a combination of a shooting method and an adaptive step size Runge-Kutta integration scheme. 

An example of the numerically obtained saddle-point solution is shown in Figs.~\ref{Figure2}(a,b). This solution corresponds to a transition between two stable solutions with respective fluxoid numbers $(n_\uparrow,n_\downarrow) = (0,1)$ and $(1,1)$. Since $n_\uparrow$ changes from 0 to 1 while $n_\downarrow$ remains the same, $f_\uparrow$ goes through a strong suppression in a region of size $\xi = \xi_0 (1-t)^{-1/2}$ centered at $\theta=0$, whereas $f_\downarrow$ is only slightly perturbed around its stable constant solution.

Once the saddle-point configurations of $\psi_\uparrow$ and $\psi_\downarrow$ are found for a transition between $(n_\uparrow,n_\downarrow)$ and $(n'_\uparrow,n'_\downarrow)$, the saddle-point free energy associated with this transition, $F_{(n_\uparrow,n_\downarrow) \leftrightarrow (n'_\uparrow,n'_\downarrow)}$, can be calculated through Eq.~(\ref{eq:FreeEnergy}). For the stable solutions with fluxoid numbers $0 \leq n_{\sigma} \leq 1$ and the saddle-point solutions connecting them, the free energies are plotted in Fig.~\ref{Figure2}(c) as a function of the field $h$. We note that there is an exact degeneracy between $(n_\uparrow,n_\downarrow) = (0,1)$ and $(1,0)$ because we neglect the Zeeman splitting by setting $b=0$ in this section.

Since the free-energy barrier is simply given by $\Delta F_{(n_\uparrow,n_\downarrow)\rightarrow (n'_\uparrow,n'_\downarrow)} = F_{(n_\uparrow,n_\downarrow)\leftrightarrow (n'_\uparrow,n'_\downarrow)}-F_{(n_\uparrow,n_\downarrow)}$ for the transition from $(n_\uparrow,n_\downarrow)$ to $(n'_\uparrow,n'_\downarrow)$, the thermally-activated rate of this transition can be estimated as~\cite{cooper_resistance_2010, halasz_fractional_2021}
\begin{equation}\label{eq:TransitionRate}
\Gamma_{(n_\uparrow,n_\downarrow)\rightarrow (n'_\uparrow,n'_\downarrow)} \propto P_{(n_\uparrow,n_\downarrow)}\text{exp}\left[{-\beta\Delta F_{(n_\uparrow,n_\downarrow)\rightarrow (n'_\uparrow,n'_\downarrow)}}\right],
\end{equation}
where $\beta = 1/T$ is the inverse temperature, and $P_{(n_\uparrow,n_\downarrow)}$ is the probability of the superconducting ring to be in the stable state $(n_\uparrow,n_\downarrow)$. At any finite temperature, this probability is given by 
\begin{eqnarray}
P_{(n_\uparrow,n_\downarrow)} &=& \frac{1}{Z}\text{exp}\left[-\beta F_{(n_\uparrow,n_\downarrow)}\right],\nonumber \\ 
Z &=& \sum_{n_\uparrow,n_\downarrow}\text{exp}\left[-\beta F_{(n_\uparrow,n_\downarrow)}\right].
\end{eqnarray}
In the presence of a small bias current $I_{\mathrm{bias}}$ applied to a section of the ring, the free-energy barrier takes the modified form~\cite{cooper_resistance_2010} $\Delta \tilde{F}_{(n_\uparrow,n_\downarrow)\rightarrow (n'_\uparrow,n'_\downarrow)} = \Delta F_{(n_\uparrow,n_\downarrow)\rightarrow (n'_\uparrow,n'_\downarrow)} - (\delta n_\uparrow + \delta n_\downarrow) \Phi_0 I_{\mathrm{bias}}/4$ with $\delta n_\sigma \equiv n'_\sigma - n_\sigma$. The mean voltage $\langle V \rangle$ between the two end points of the section is then proportional to
\begin{widetext}
\begin{equation}
\langle V \rangle \propto \sum_{n_\uparrow,n_\downarrow} P_{(n_\uparrow,n_\downarrow)} \sum_{n'_\uparrow,n'_\downarrow} \left( \delta n_\uparrow + \delta n_\downarrow \right) \text{exp} \left[ -\beta \Delta F_{(n_\uparrow,n_\downarrow)\rightarrow (n'_\uparrow,n'_\downarrow)} + \beta \left( \delta n_\uparrow + \delta n_\downarrow \right) \frac{ \Phi_0 I_{\mathrm{bias}}} {4} \right].
\end{equation}
\end{widetext}
Finally, if we assume $I_{\mathrm{bias}} \ll T / \Phi_0$, the effective resistance takes the form
\begin{widetext}
\begin{equation}\label{eq:Resistance}
R_{\mathrm{eff}} = \frac{\langle V \rangle} {I_{\mathrm{bias}}} \propto \frac{\sum\limits_{n_\uparrow,n_\downarrow}\left\{\sum\limits_{\delta n_\uparrow = \pm 1}\delta n^2_\uparrow \text{exp} \left[ -\beta F_{(n_\uparrow,n_\downarrow) \leftrightarrow (n_\uparrow + \delta n_\uparrow, n_\downarrow)} \right] + \sum\limits_{\delta n_\downarrow = \pm 1}\delta n^2_\downarrow \text{exp} \left[ -\beta F_{(n_\uparrow,n_\downarrow) \leftrightarrow (n_\uparrow, n_\downarrow + \delta n_\downarrow)} \right] \right\}} {\sum\limits_{n_\uparrow,n_\downarrow} \text{exp} \left[ -\beta F_{(n_\uparrow,n_\downarrow)} \right]}.
\end{equation}
\end{widetext}
Here we ignore transitions that involve both fluxoid numbers $n_\sigma$ or change either fluxoid number by more than $1$ as they have larger free-energy barriers and their contributions to the resistance are negligible. We also disregard overall temperature-dependent prefactors from both the transition rate~\cite{mccumber_time_1970} and the expansion of $\langle V \rangle$ up to first order in $I_{\mathrm{bias}}$ as they only depend on the temperature via power laws, i.e., much weaker than the exponentials in Eq.~(\ref{eq:Resistance}).

\begin{figure*}
\centering
\includegraphics[scale = 0.3]{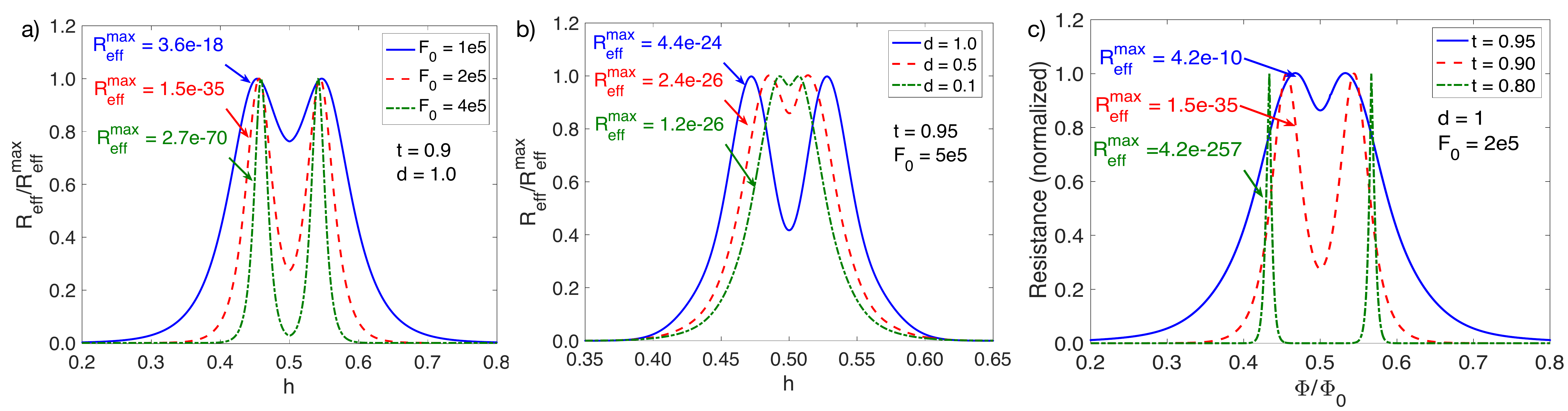}
\caption{Magnetoresistance oscillations of a spin-triplet superconducting ring for different values of (a) the dimensionless energy parameter (i.e., condensation energy) $F_0$, (b) the coupling coefficient $d$, and (c) the dimensionless temperature $t$. The following parameter values apply to all three subfigures: $r_0 = 20$, $c = 0.1$, and $b = 0$. The resistance $R_{\mathrm{eff}}$ is calculated from Eq.~(\ref{eq:Resistance}) and normalized by its maximum value, $R_{\mathrm{eff}}^{\mathrm{max}}$, which is specified for each resistance curve.}
\label{Figure3}
\end{figure*}

Figure~\ref{Figure3} shows the magnetoresistance calculated from Eq.~(\ref{eq:Resistance}) at different values of the temperature $t$, the coupling constant $d$, and the energy parameter $F_0$. We readily observe the expected two-peak structure in the magnetoresistance oscillations~\cite{vakaryuk_effect_2011, yasui_little-parks_2017, cai_magnetoresistance_2022} and recognize clear trends in its behavior as a function of the three parameters. We first consider the separation between the two peaks that depends on $t$ and $d$ but not on $F_0$. As understood in previous works~\cite{vakaryuk_effect_2011, yasui_little-parks_2017, cai_magnetoresistance_2022}, each peak corresponds to a transition between a FQV state and a HQV state, with a HQV state being energetically favored in the narrow range between the two peaks. Hence, we immediately identify the separation between the two peaks as the width parameter $\Delta h$ in Eq.~(\ref{eq:Delta-1}) and understand that, for $d (1-t) \ll 1$, it is linearly proportional to both the coupling constant $d$ and the temperature difference $1-t$ with respect to the critical temperature.

We next focus on the width of each peak which, in conjunction with the peak separation, determines how well the two peaks are distinguishable from each other. In general, we observe that the peaks broaden if $F_0$ is decreased or $t$ is increased. To understand this result, we first notice from Eq.~(\ref{eq:Resistance}) that each peak extends over a field range in which the free-energy difference between the two lowest-energy fluxoid states $(n_\uparrow,n_\downarrow)$ is smaller than the temperature $T$; the resistance has a peak within such a field range because thermal phase slips between the two highly occupied lowest-energy states occur with an enhanced rate. Then, using Eq.~(\ref{eq:FreeEnergy_minima}) and assuming $r_0 \gg 1$ as well as $|1-t| \ll 1$, we find $|\partial F_{(n_\uparrow,n_\downarrow)} / \partial h| \sim F_0 (1-t) r_0^{-2}$ and estimate the peak width as $\delta h \sim T / |\partial F_{(n_\uparrow,n_\downarrow)} / \partial h| \sim r_0^2 T_c / [F_0 (1-t)]$. The conclusion is that the peak width is inversely proportional to both the energy parameter $F_0$ and the temperature difference $1-t$ with respect to the critical temperature. We note that the energy parameter $F_0$ is proportional to the total volume of the superconducting ring and can, in principle, be made arbitrarily large by increasing the out-of-plane ring height $L$; see the next section for quantitative estimates.

Even if the peaks are well separated and sufficiently narrow, the two-peak structure may still not be observable if the overall magnitude of the resistance is prohibitively small. According to Fig.~\ref{Figure3}, this problem occurs when $F_0$ is too large and/or $t$ is too small. From Eq.~(\ref{eq:Resistance}), the overall magnitude of the resistance is determined by the free-energy difference between the relevant stable and saddle-point solutions (i.e., the free-energy barrier). Since the saddle-point solutions have a localized suppression of superconductivity in a region of size $\xi = \xi_0 (1-t)^{-1/2}$ with respect to the stable solutions, this free-energy difference can be estimated as $\delta F \sim |F_{(n_\uparrow,n_\downarrow)}| \xi / R_0 \sim F_0 (1-t)^{3/2} r_0^{-1}$ if we assume $r_0 \gg 1$ and $|1-t| \ll 1$ again. The overall magnitude of the resistance is then found to be prohibitively small when $\beta \delta F \sim F_0 (1-t)^{3/2} / (r_0 T_c) \sim r_0 (1-t)^{1/2} / \delta h \gg 1$. Interestingly, this result reveals a fundamental trade-off between increasing the magnitude of the resistance and decreasing the peak width, which are both important for observing the two-peak structure. Still, Fig.~\ref{Figure3} demonstrates that a discernible two-peak structure with a reasonable magnitude of the resistance is possible with the right parameter values.

\section{Discussion and outlook}

In this work, we employed the Ginzburg-Landau approach to study the Little-Parks magnetoresistance oscillations of spin-triplet superconductors in the presence of HQVs harboring Majorana bound states. Focusing on a ring geometry with a sufficiently small width ($W \ll \xi, \lambda$), we first constructed the appropriate Ginzburg-Landau free energy and identified a specific higher-order term [the one proportional to $d$ in Eq.~(\ref{eq:FreeEnergy})] that is critical for stabilizing HQVs. Then, we used the associated Ginzburg-Landau equations to derive both the ``conventional'' Little-Parks oscillations of the superconducting critical temperature~\cite{little_observation_1962} and the closely related oscillations in the residual resistance due to thermal vortex tunneling below the critical temperature~\cite{cai_magnetoresistance_2022, sochnikov_large_2010, sochnikov_oscillatory_2010, cai_unconventional_2013, mills_vortex_2015}.

Our main result is a rigorous theoretical underpinning of the characteristic two-peak structure that is expected in the magnetoresistance oscillations~\cite{vakaryuk_effect_2011, yasui_little-parks_2017, cai_magnetoresistance_2022}. The two peaks demarcate a narrow field range of width $\Delta h$ [see Eq.~(\ref{eq:Delta-1})] in which it is energetically favorable to bind a HQV to the central hole of the superconducting ring. Since HQVs are stabilized by a higher-order term in the Ginzburg-Landau free energy, the peak separation $\Delta h$ is linearly proportional not only to the appropriate coupling constant $d$ but also to the relative temperature $1-t$ with respect to the critical temperature. In other words, the two-peak structure is observable in the magnetoresistance oscillations below the critical temperature but not in the ``conventional'' Little-Parks oscillations of the critical temperature itself.

We also established the general conditions under which the two-peak structure in the magnetoresistance oscillations is experimentally discernible. In particular, we identified a fundamental trade-off between making the peaks sufficiently narrow (hence, distinguishable from each other) and ensuring that the overall magnitude of the resistance is not overly suppressed. For the parameter values $r_0 = R_0 / \xi_0 \sim 10$ and $t = T / T_c \sim 0.9$ used in Fig.~\ref{Figure3}, we find that this trade-off corresponds to energy parameter $F_0 \sim 10^5$. Since $T_c F_0$ is defined as a total condensation energy at zero temperature, this quantity is on the order of $\nu \Delta^2 \sim \nu T_c^2 $, where $\nu$ is the electronic density of states at the Fermi level, and $\Delta$ is the superconducting pairing gap. Thus, we readily obtain $F_0 \sim \nu T_c \sim V_0 T_c / (a^3 E_F)$, where $V_0$ is the ring volume, $a$ is the lattice constant, and $E_F$ is the Fermi energy. For $T_c \sim 1$ K, $a \sim 1$ nm, and $E_F \sim 0.1$ eV, the ideal energy parameter $F_0 \sim 10^5$ then corresponds to ring volume $V_0 \sim 0.1$ $\mu$m$^3$, which is achieved, for example, by setting the ring radius, the ring width, and the ring height to $R_0 \sim 1$ $\mu$m, $W \sim 100$ nm, and $L \sim 100$ nm, respectively.

While the critical temperature marking the onset of superconductivity (i.e., the upper critical temperature) does not have a distinctive structure in its Little-Parks oscillations, we also found a lower critical temperature separating a single-spin superconducting state above and a spin-triplet superconducting state below. In turn, this lower critical temperature has two distinct minima around each half-integer value of $h$ [see Fig.~\ref{Figure1}(c)], which correspond to the two peaks in the related magnetoresistance oscillations~\footnote{Since the resistance decreases if the critical temperature is increased, it is not surprising that a two-peak structure in the resistance translates into a \emph{two-valley} structure in the critical temperature.}. To experimentally determine the lower critical temperature, one would need to selectively measure the electrical resistance of the spin species that is normal above and superconducting below the transition. For example, one could apply the bias current through half-metallic leads that can only emit or absorb one spin species but not the other one. We further note that this measurement would require a sharp superconducting transition with no residual resistance due to vortex tunneling below. As such, it would be in a different regime and need a much larger ring volume $V_0$ than discussed above. Since the radius and the width are restricted to $R_0 \lesssim 1$ $\mu$m and $W \lesssim 100$ nm by other considerations, this measurement would then correspond to a cylindrical geometry with a large height $L \gg 1$ $\mu$m as in the original Little-Parks experiment~\cite{little_observation_1962}.

Physically, the higher-order term proportional to $d$ in the Ginzburg-Landau free energy stabilizes HQVs by penalizing the charge supercurrent but not the spin supercurrent. In this sense, it plays an analogous role to the ratio of the spin and charge superfluid densities, $\gamma = \rho_{\mathrm{sp}} / \rho_{\mathrm{s}}$, in the London limit~\cite{chung_stability_2007, halasz_fractional_2021}. Indeed, from a comparison of Eq.~(\ref{eq:FreeEnergy_minima}) in this work and Eq.~(15) in Ref.~\onlinecite{halasz_fractional_2021}, we can identify the superfluid-density ratio as $\gamma = (1+c) / [(1+c) + 2(1-t)d]$, which is consistent with the general expectation that $\gamma \to 1$ at the critical temperature~\cite{vakaryuk_effect_2011}. Since it is also expected on general grounds that $\gamma < 1$ for interacting superconductors~\cite{chung_stability_2007, leggett_inequalities_1968, leggett_theoretical_1975}, we anticipate that, as a result of $d > 0$, our results apply to any spin-triplet superconductor in which $(\uparrow\uparrow)$ and $(\downarrow\downarrow)$ Cooper pairs are energetically favored over $(\uparrow\downarrow) + (\downarrow\uparrow)$ Cooper pairs. In the future, it would be interesting to understand how our results connect to the London limit and, in particular, how the fractional magnetoresistance oscillations in the presence of disorder predicted in Ref.~\onlinecite{halasz_fractional_2021} can be recovered from our general Ginzburg-Landau approach. This connection could then be used to establish the general conditions under which the fractional magnetoresistance oscillations are experimentally observable.

\section*{Acknowledgments}
This work was supported by the U. S. Department of Energy, Office of Science, Basic Energy Sciences, Materials Sciences and Engineering Division. This manuscript has been authored by employees of UT-Battelle, LLC under Contract No.~DE-AC05-00OR22725 with the U.S. Department of Energy. The U.S. Government retains and the publisher, by accepting the article for publication, acknowledges that the U.S. Government retains a nonexclusive, paid-up, irrevocable, worldwide license to publish or reproduce the published form of this manuscript, or allow others to do so, for U.S. Government purposes.

\end{document}